\documentclass[EN]{rcftex}
\usepackage{graphicx}
\usepackage{amssymb}
\usepackage{amsfonts}
\usepackage{amsmath}
\usepackage[T1]{fontenc}

\title{Terahertz absorption by excitonic polaritons}

\author{Carlos Andr\'es Vera-Ciro$^{a}$, Alain Delgado$^{b}$, and Augusto Gonz\'alez$^{c}$}

\affiliations{$^{a}$ Kapteyn Astronomical Institute, University of Groningen, P.O. Box 800, 9700 AV Groningen, The Netherlands\\
$^{b}$Centro de Aplicaciones Tecnol\'ogicas y Desarrollo
Nuclear, Calle 30 No 502, Miramar, CP 11300, La Habana, Cuba\\
$^{c}$Instituto de Cibern\'etica, Matem\'atica y F\'isica, Calle E 309, Vedado, CP 10400,
 La Habana, Cuba\emailto}

\abstractES{La teor\'ia de respuesta lineal es utilizada para calcular el espectro de absorci\'on,
en la banda de terahertz, de un sistema polarit\'onico compuesto por excitones en un punto cu\'antico 
fuertemente
acoplados al modo fot\'onico fundamental confinado en un micropilar. En un sistema termalizado
(condensado de Bose) la funci\'on espectral muestra un pico, asociado a la transici\'on 
excit\'onica $1s - 2p$, reforzada por efectos polarit\'onicos. Por el contrario, en un sistema no
equilibrado el pico de absorci\'on se localiza a bajas energ\'ias. Luego, la medici\'on de la
absorci\'on de terahertz podr\'ia indicar el grado de termalizaci\'on de los polaritones.}

\abstractEN{
We use linear response theory in order to compute the light absorption spectrum, in the terahertz band, 
of a polariton system composed by excitons in a quantum dot very strongly coupled to the
lowest photon mode of a thin micropillar. In a thermalized (Bose condensed) system at
low temperatures, the spectral function shows a peak associated to a $1s - 2p$ like 
exciton transition, enhanced by polariton effects. On the other hand, in a non-equilibrium 
system absorption is peaked at low energies. Thus, a measurement of terahertz absorption 
could give an indication of the degree of thermalization in the polariton system.}

\keyW{Terahertz absorption, Excitonic polaritons}

\begin{document}
\maketitle

The strong coupling regime in the interaction between a confined
photon mode and electron-hole pairs in semiconductor nanodevices has
been demonstrated recently \cite{strongcoupling}. The
quasiparticles, so called polaritons \cite{Hopfield,polaritons}, which are
roughly half excitons and half photons, offer very interesting
possibilities, such as, for example, a new lasing mechanism (polariton lasing) based on
their quasibosonic nature \cite{polaritonlasing}, with pumping threshold 
(related to ground-state occupation) two orders of
magnitude lower than ordinary (photon) lasing in the same devices
\cite{Thresholds}, and operation at ambient temperatures
\cite{300K}.

In the present paper, we focus on the linear response of a model
polariton system to terahertz radiation. The first motivation to carry
on such a study is the intuitive idea that the
interaction with the confined photon mode reinforces coherence of
the excitonic subsystem and, thus, may reinforce the collective
response of the excitons to the terahertz probe. This may result in a
semiconductor version of the Giant Dipole Resonances (GDR), a phenomenon
widely studied in nuclei \cite{GDRnuclei} and electron clusters
\cite{GDRclusters}, with the possibility of controlling the position
and intensity of the resonance by varying parameters such as the
pumping rate or the photon-exciton detuning. 

The second good reason to study terahertz absortion by excitonic polaritons
is that it has proven to be very useful in order to observe exciton
formation dynamics in quantum wells \cite{Chemla}, and bulk systems 
\cite{Shimano}. In the polariton system, a few years ago the common belief
was that a thermalized Bose-condensed state is reached \cite{LesiDang,Yamamoto}.
Very recently, however, this conclusion along with the interpretation of
most experiments is being questioned \cite{Kavokin}.
We think, the available experimental techniques should
be able to measure the degree of thermalization of the polariton system,
not only under stationary conditions \cite{LesiDang}, but in the pumped
regime as well \cite{Yamamoto}. Indicators following from interband emission 
alone are not enough because the main qualitative features (population of the
lowest polariton state, behavior of the second order coherence function, etc) 
can be reproduced also from dynamical equations, without any thermalization 
mechanisms, both in the pumped \cite{nuestraPRL} and in the stationary regimes 
\cite{PRBnuestra}.

\begin{figure}
\includegraphics[width=8cm]{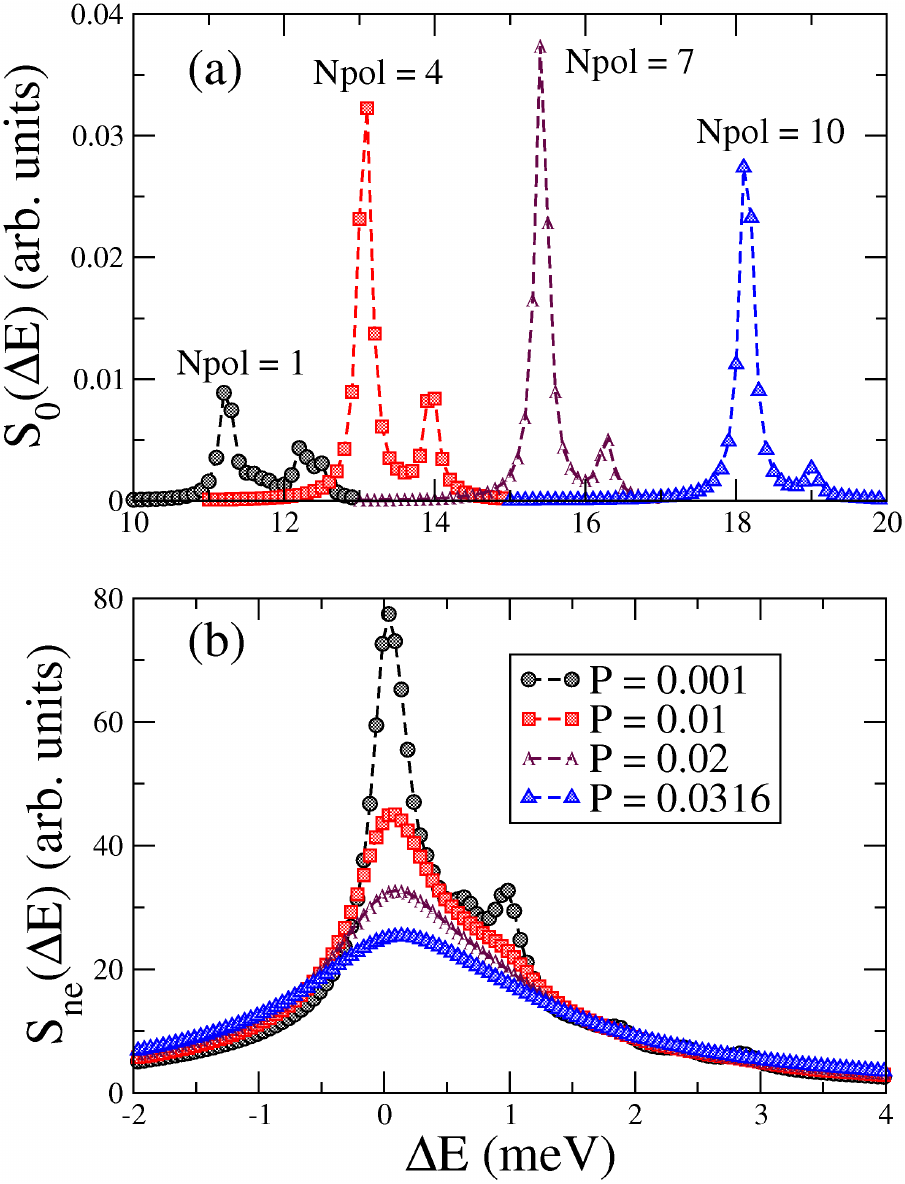}
\caption{\label{fig1} (Color online) Comparison between the equilibrium and
non-equilibrium terahertz absorption. (a) Ground-state spectral function, Eq. 
(\ref{eq1}), for various $N_{pol}$ numbers. At a given $N_{pol}$, the GDR is
the highest peak. (b) Non-equilibrium
spectral function, Eq. (\ref{eq12}), for pumping rates (in ps$^{-1}$) 
corresponding to mean polariton number in the interval (1,10). 
The detuning parameter is $\Delta=-3$ meV.}
\end{figure}

Below, we compute terahertz absorption in two extreme situations.
One is a Bose condensed state at very low temperatures, in such a
way that only the many-particle ground state has a significant occupation 
probability. In our model, with not very realistic parameters, the $1s - 2p$ 
excitonic transition is located at around 10 meV, that is the temperature 
should be lower than 100 K, a common experimental situation. The spectral 
function shows a GDR-like peak, whose position grows with the polariton number, 
Fig. \ref{fig1}a).  

The second case corresponds to a polariton system in a non-equilibrium 
stationary state (result of a balance between pumping and losses),
with occupation probabilities that can not be described by a Gibbs 
distribution. The terahertz spectral function gets a completely
different shape, with a central peak at near zero energy which practically
does not depend on the pumping rate, Fig. \ref{fig1}b). 

Intermediate, real experimental, situations would interpolate between the 
two extremes, and a measurement of the response in real systems would indicate 
their degree of thermalization. 

Calculations are carried on in a model for the quantum dot - microcavity
system, detailed described in Ref. [15], with very 
strong light-matter coupling constant (3 meV), which leads to a 
significant blueshift of the GDR resonance with respect to the $1s - 2p$ like
exciton transition. The main qualitative conclusions of the paper
are expected to be valid also for any relatively large quantum dot or
thin quantum well micropillar working under the strong coupling regime. 

{\bf (i) Ground-state response of non-interacting polaritons}

In order to get a preliminary estimate of the absorption spectrum,
we first consider the ground-state response of non-interacting 
polaritons. We assume the system is in a Bose-condensed state, 
with $N_{pol}$ polaritons occupying a single state. Intraband absorption
is described by the dipole operator acting only on the exciton functions.
The absorption probability is then proportional to 
$|\alpha~d_{10}|^2 N_{pol}$, where $\alpha$ is the Hopfield coefficient
\cite{Hopfield} (that is, the weight of the exciton in the 
polariton function), and $d_{10}$ is the
intra-band dipole matrix element between ground-state exciton and 
an excited-state function. The latter is supposed to concentrate the oscillator
strength for dipole transitions.  Notice that the absorption probability
increases with the number of polaritons in the ground state.
The peak position, on the other hand, should be almost constant, 
roughly equal to the energy difference between the exciton ground- and 
excited states.

Finite, but low, temperatures, should lead to similar results.
In a grand canonical description, on the other hand, which is
more natural for the polariton system, sectors with polariton number
near the mean value will contribute also to the spectral function
with relatively high weights. The effects of polariton-polariton 
interactions is considered in the next paragraph.

{\bf (ii) Ground-state response of interacting polaritons}

Polariton-polariton interactions come from residual Coulomb 
interactions between excitons. Instead of using a phenomenological
approach, we start from a model in which Coulomb interactions 
are treated exactly, and the fermionic degrees of freedom are
explicit. There is a finite number (10) of single-particle states 
for electron and holes, and a single photon mode. Saturation effects
due to Fermi statistics are seen when the polariton number is around
(or greater than) 10. A detailed description can be found elsewhere 
\cite{PRBnuestra}.

\begin{figure}
\includegraphics[width=8cm]{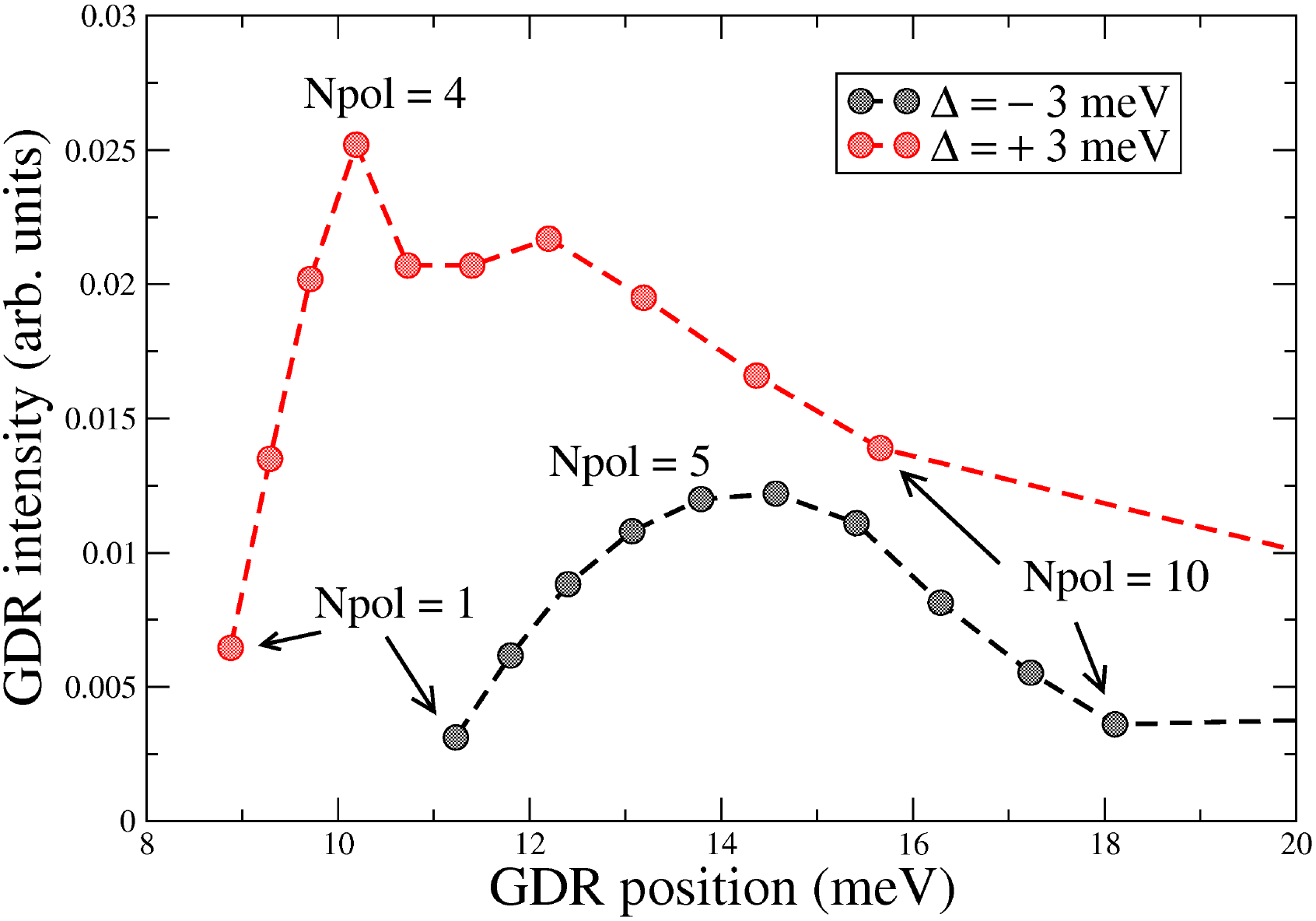}
\caption{\label{fig2} (Color online) Intensity (that is, dipole matrix
elements squared) and position of the GDR peak
for two different values of the detuning, $\Delta$. Each dot
corresponds to a given $N_{pol}$.}
\end{figure}

The very-low temperature (ground-state) response of the $N_{pol}$-polariton 
system is contained in the spectral function:

\begin{equation}
S_0(\omega)=\sum_I \frac{|\langle I|d|J\rangle|^2 \Gamma_0/\pi}
{\Gamma_0^2+(\omega_{IJ}-\omega)^2},
\label{eq1}
\end{equation}

\noindent
where matrix elements, $\langle I|d|J\rangle$, of the
intraband dipole operator, $d\sim \sum_i (\vec r_i^{(h)}-\vec
r_i^{(e)})$ (where $\vec r^{(h)}$ and $\vec r^{(e)}$ are,
respectively, the hole and electron position vectors) shall be
computed. $|J\rangle$ is the ground state function of the $N_{pol}$-polariton 
system, and $|I\rangle$ are excited states. $\Gamma_0=0.1$ meV/$\hbar$
is a phenomenological damping parameter, and $\omega_{IJ}=(E_I-E_J)/\hbar$
-- the transition frequencies.

In our model, wave 
functions are constructed as linear combinations:

\begin{equation}
|P\rangle=\sum_{S_e,S_h,n} C_{S_e,S_h,n}|S_e,S_h,n\rangle,
\label{eq2}
\end{equation}

\noindent where $S_e$ and $S_h$ are Slater determinants for
electrons and holes, with electron and hole numbers $N_e$ and $N_h$,
respectively, and $n$ is the number of photons in the confined mode.
Functions entering the combination preserve the polariton number:

\begin{equation}
N_{pol}=N_e+n=N_h+n,
\label{eq3}
\end{equation}

\noindent
and the total (envelope) angular momentum projection
along the cavity axis (we assume a circular section):

\begin{equation}
L=\sum_i (l^{(e)}_i+l^{(h)}_i).
\label{eq4}
\end{equation}

\noindent
In Eq. (\ref{eq4}), the index $i$ labels the particles. $l^{(e)}_i$, for example, 
corresponds to the angular momentum projection
along the cavity axis of the $i$-th electron. The ground-state function, 
$|J\rangle$, has $L=0$, whereas $|I\rangle$ are $L=1$ functions.

We show in Fig. \ref{fig1} (a) the spectral function for 
different polariton numbers and detuning $\Delta=-3$ meV. In the
model, the parameter $\Delta$ measures the photon energy with
respect to the nominal band gap, not the photon-exciton detuning.
$\Delta=-3$ meV approximately corresponds to resonant conditions.

The GDRs can be identified as the dominant peaks in these curves.
The peak position monotonously increases with increasing
polariton number. This can be understood on intuitive grounds.
The mass of the electron (or hole) cloud is $m\sim N_{pairs}$,
and the Hooke coefficient for the force acting between clouds is 
$k\sim N_{pairs}^2$. Then, the excitation energy of the dipole
mode is $\hbar\omega\sim \sqrt{k/m}\sim\sqrt{N_{pairs}}\sim\sqrt{N_{pol}}$.
The maximum intensity, on the other hand, has a non-trivial dependence 
on $N_{pol}$, a kind of saturation effect is observed.
The intensity first increases, as in the non-interacting case, 
but then, after reaching a maximum value, decays. 
These dependences are illustrated in Fig. \ref{fig2}, where
the case $\Delta=+3$ meV, corresponding to an enhanced excitonic
component of polaritons, is also shown. In this positive detuning
situation, the absorption probability rises because the Hopfield
parameter $\alpha$ increases.

In spite of the fact that calculations are performed in a particular model, 
we expect that the statement about the existence of a peak
in the absorption spectrum at relatively high excitation energies (of the order 
of the exciton $1s - 2p$ transition), whose intensity increases at least for
polariton numbers well below saturation values, is general enough, and could be 
used as a criterium of a low-temperature system in an , equilibrium (Bose-condensed)
stae.

{\bf (iii) Dynamical response of the non-equilibrium system (with
non-resonant pumping and photon losses)}

Below, we assume that relaxation mechanisms are not effective, and 
can not lead the polariton system to an equilibrium (thermal) state. 
The system is, thus, described
by a density matrix, which is obtained from a master equation
that takes care of photon losses through the cavity mirrors and
incoherent (non-resonant) pumping. Details can be found in Ref.
[15]. We solve the master equation in the
stationary ($t\to\infty$) limit in order to obtain the quasiequilibrium
distribution, $\rho^{(\infty)}$.

The response to the terahertz probe is computed in the linear
approximation, where the probe does not modify the
quasiequilibrium distribution. We adopt a computational scheme
similar to the one used for the photoluminescence response
\cite{PRBnuestra}. The starting point is the first-order correlation
function:

\begin{equation}
\langle d^{\dagger}(t+\tau)d(t)\rangle=\sum_{I,J}\langle
J|d^{\dagger}|I\rangle g_{d,IJ}, \label{eq5}
\end{equation}

\noindent written in terms of the auxiliary function:

\begin{equation}
g_{d,IJ}(t+\tau,t)=\langle(|J\rangle\langle I|)(t+\tau)d(t)\rangle,
\label{eq6}
\end{equation}

\noindent
where $|J\rangle$ are $N_{pol}$-polariton functions with total
angular momentum $L=0$, and the $|I\rangle$ are
$N_{pol}$-polariton functions with $L=1$. Because of the Quantum
Regression Theorem \cite{QRT}, $g_{d,IJ}$ satisfies the
same equation as the density matrix, that is \cite{PRBnuestra}:

\begin{eqnarray}
\frac{\rm d}{\rm d\tau}g_{d,IJ}&=&(i\omega_{IJ}-\Gamma_{IJ})
 g_{d,IJ}\nonumber\\
 &+& \kappa\sum_{K,M}\langle I|a|M\rangle g_{d,MK}
  \langle K|a^{\dagger}|J\rangle\nonumber\\
 &-&\frac{\kappa}{2}\sum_{K\ne I,M}\langle I|a^{\dagger}|M\rangle
  \langle M|a|K\rangle g_{d,KJ}\nonumber\\ 
 &-&\frac{\kappa}{2}\sum_{K,M\ne J}g_{d,IM}
 \langle M|a^{\dagger}|K\rangle\langle K|a|J\rangle, \label{eq7}
\end{eqnarray}

\noindent with boundary conditions at $t\to\infty$, $\tau=0$:

\begin{eqnarray}
g_{d,IJ}&=&\sum_K\langle I|d|K\rangle \rho_{KJ}^{(\infty)}\nonumber\\
 &\approx& \langle I|d|J\rangle \rho_{JJ}^{(\infty)}, \label{eq8}
\end{eqnarray}

\noindent where, in the last step, we used the fact that
$\rho_{KJ}^{(\infty)}$ is approximately diagonal in the energy
representation \cite{nuestraPRL}.

In Eq. (\ref{eq7}), $\kappa$ is
the loss rate, 0.1 ps$^{-1}$ in our model. The widths,
$\Gamma_{IJ}$, are computed from:

\begin{eqnarray}
\Gamma_{IJ}&=&\frac{\kappa}{2}\sum_K\left\{ \right|\langle K|a|I\rangle|^2
 +|\langle K|a|J\rangle|^2\}\nonumber\\
 &+&\frac{P}{2}\{N_{up}(I)+N_{up}(J)\}, \label{eq9}
\end{eqnarray}

\noindent where $P$ is the pumping rate, and $N_{up}(I)$ is the
number of states with polariton number $N_{pol}(I)+1$ used to solve
the equations.

The general solution of the linear system, Eq. (\ref{eq7}), is
written in terms of the eigenvalues, $\lambda_n$, and eigenvectors,
$X_{IJ}^{(n)}$, of the matrix $B_{IJ,MK}$ defined by the r.h.s. of
Eq. (\ref{eq7}), that is:

\begin{equation}
g_{d,IJ}(\tau)=\sum_n C_n \exp{(\lambda_n\tau)} X_{IJ}^{(n)},
\label{eq10}
\end{equation}

\noindent where the coefficients $C_n$ are determined from the
boundary conditions, Eq. (\ref{eq8}).

The Fourier transform of Eq. (\ref{eq5}) defines the response
spectral function to the terahertz probe in the quasi-equilibrium
system:

\begin{equation}
S_{ne}(\omega)=-\frac{1}{\pi} \sum_{I,J}\sum_n\frac
{D^{(r)}_{IJ,n}\lambda_n^{(r)}+D^{(i)}_{IJ,n}(\lambda_n^{(i)}-\omega)}
{(\lambda_n^{(r)})^2+(\lambda_n^{(i)}-\omega)^2}, \label{eq11}
\end{equation}

\noindent where $D_{IJ,n}=\langle J|d^{\dagger}|I\rangle C_n
X_{IJ}^{(n)}$, and superscripts $(r)$, $(i)$ refer to the real and imaginary parts
of the magnitudes, respectively.

A simplified and more intuitive expression comes from the diagonal
terms of Eq. (\ref{eq7}).\cite{PRBnuestra} Notice that, for excitation energies 
$\hbar\omega > 1$ meV, the diagonal is at least 10 times higher
than the off-diagonal elements (because $\kappa=0.1$ ps$^{-1}$).
Neglecting the off-diagonal terms, we get:

\begin{equation}
S_{ne}(\omega)\approx\frac{1}{\pi} \sum_{I,J}\frac
{|\langle I|d|J\rangle|^2\rho_{JJ}^{(\infty)}\Gamma_{IJ}}
{\Gamma_{IJ}^2+(\omega_{IJ}-\omega)^2}.
\label{eq12}
\end{equation} 

\begin{figure}
\includegraphics[width=8cm]{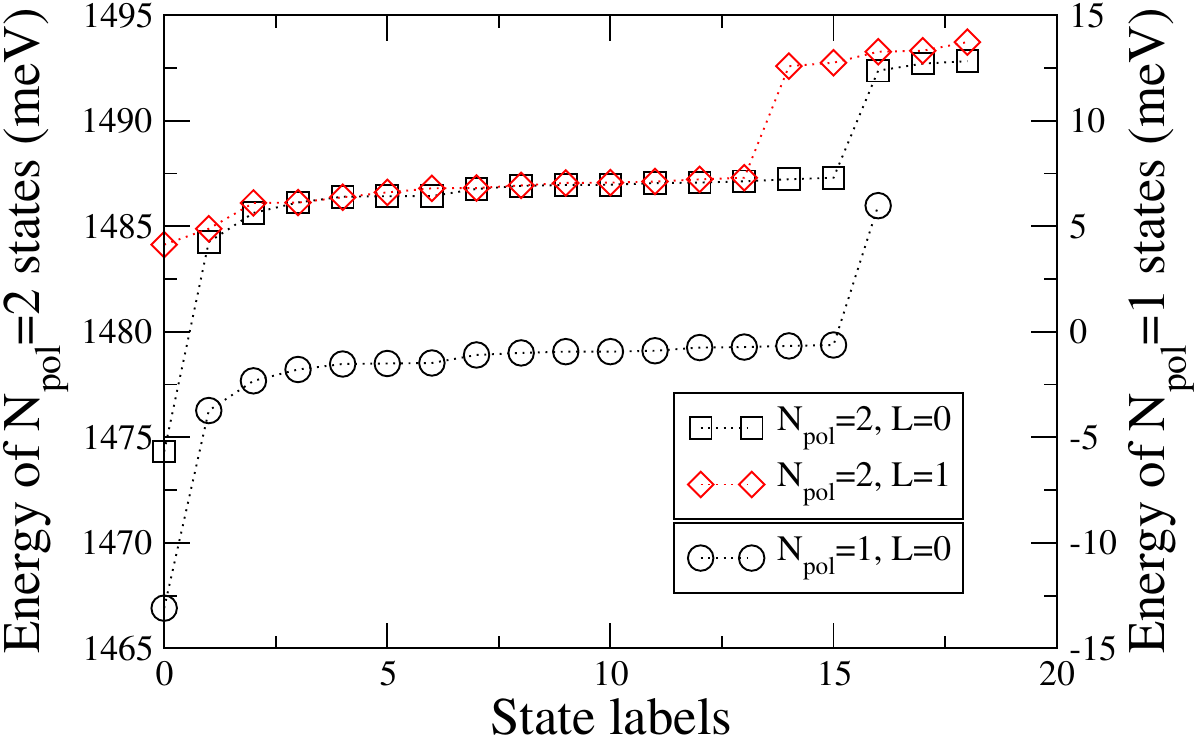}
\caption{\label{fig3} (Color online) The lowest $N_{pol}=2$ states with
$L=0$ and $L=1$ in the model. A big number of near zero-energy dipole 
transitions are possible in the $N_{pol}=2$ sector. We draw in the same figure,
shifted by the nominal $E_{gap}$, the $N_{pol}=1$, $L=0$ states. Notice that 
$L=0$ bands with different $N_{pol}$ numbers are almost parallel.}
\end{figure}

As compared with $S_0$, the non-equilibrium spectral function
includes also contributions from the excited states, $|J\rangle$, 
which may have relatively high occupation probabilities,
$\rho_{JJ}^{(\infty)}$, as can be seen, for example, in Fig.
6 of Ref. [15]. On the other hand, the 
dipole matrix elements for transitions originated in excited
states could be much stronger than ground-state dipole elements. This 
statement follows from the energy-weighted sum rule for dipole
transitions \cite{Ring,PhysEnuestra}:

\begin{equation}
\sum_I \Delta E_{IJ} ~|\langle I|d|J\rangle|^2=C,
\label{eq13}
\end{equation}

\noindent
where constant $C$ does not depend on the indices $J$.

The sum in Eq. (\ref{eq13}) reduces to a single term when the oscillator strength
from state $|J\rangle$ is concentrated on a single state, $|I\rangle$. Then, if
there were excited states $|J\rangle$ for which the dominant transitions
have $\Delta E_{IJ}\sim 0.1$ meV, for example, 
their contribution to $S_{ne}$ would be 100 times stronger than the ground state
contribution. This is, indeed, what one sees in the spectral
function, Fig. \ref{fig1} (b). An extra factor of around 20
comes from the number of excited states. We have drawn in this
picture the non-equilibrium spectral function for pumping rates,
$P$, corresponding approximately to the same situations depicted
in Fig. \ref{fig1} (a). That is, the mean polariton number 
($\langle N_{pol}\rangle=\sum_J \rho_{JJ}^{(\infty)}N_{pol}(J)$) for $P=0.01$
ps$^{-1}$, for example, is around 4, etc. In Fig. \ref{fig3}, we
show that near zero-energy dipole transitions are very common in our model,
and should be very common also in micropillars with embeebed quantum wells
because of the exciton near flat band.

In conclusion, we expect the absorption spectral function for a
non-equilibrium polariton system to be peaked at near zero energies,
in clear contrast with the Bose-condensed system, whose spectral
function is peaked at the GDR. The dependence on $N_{pol}$ is
also very different. In the thermalized system absorption increases
with increasing polariton number, whereas in the nonequilibrium system
it decreases as the pumping rate increases. Thus, terahertz absorption
could be a sharp criterium allowing to discriminate between the
thermalized and the non-equilibrium scenarios. 

\acknow{A.D. and A.G. acknowledge the Caribbean Network for Quantum Mechanics, 
Particles and Fields (ICTP) for support. The authors would like to thank 
Herbert Vinck-Posada and Alexey Kavokin for discussions and help.}

\end{document}